\documentclass[3p, twocolumn]{elsarticle}
\usepackage{graphicx}
\usepackage{epsfig}
\usepackage{amssymb}
\usepackage{amsmath}

\begin{document}

\begin{frontmatter}

\title{Inelastic Neutron Scattering Studies of the Spin and Lattice Dynamics in Iron Arsenide Compounds}
\author[ANL]{R. Osborn}
\ead{rosborn@anl.gov}
\author[ANL]{S. Rosenkranz}
\author[ANL,ISIS]{E. A. Goremychkin}
\author[ORNL]{A. D. Christianson}

\address[ANL]{Materials Science Division, Argonne National Laboratory, Argonne, Illinois 60439, USA}
\address[ISIS]{ISIS Pulsed Neutron and Muon Source, Rutherford Appleton Laboratory, Didcot OX11 0QX, UK}
\address[ORNL]{Neutron Scattering Science Division, Oak Ridge Naitonal Laboratory, Oak Ridge, TN 37831, USA}

\begin{abstract} 
Although neutrons do not couple directly to the superconducting order parameter, they have nevertheless played an important role in advancing our understanding of the pairing mechanism and the symmetry of the superconducting energy gap in the iron arsenide compounds. Measurements of the spin and lattice dynamics have been performed on non-superconducting `parent' compounds based on the LaFeAsO (`1111') and BaFe$_2$As$_2$ (`122') crystal structures, and on electron and hole-doped superconducting compounds, using both polycrystalline and single crystal samples. Neutron measurements of the phonon density-of-state, subsequently supported by single crystal inelastic x-ray scattering, are in good agreement with \textit{ab initio} calculations, provided the magnetism of the iron atoms is taken into account. However, when combined with estimates of the electron-phonon coupling, the predicted superconducting transition temperatures are less than 1\,K, making a conventional phononic mechanism for superconductivity highly unlikely. Measurements of the spin dynamics within the spin density wave phase of the parent compounds show evidence of strongly dispersive spin waves with exchange interactions consistent with the observed magnetic order and a large anisotropy gap. Antiferromagnetic fluctuations persist in the normal phase of the superconducting compounds, but they are more diffuse. Below T$_c$, there is evidence in three `122' compounds that these fluctuations condense into a resonant spin excitation at the antiferromagnetic wavevector with an energy that scales with T$_c$. Such resonances have been observed in the high-T$_c$ copper oxides and a number of heavy fermion superconductors, where they are considered to be evidence of $d$-wave symmetry. In the iron arsenides, they also provide evidence of unconventional superconductivity, but a comparison with ARPES and other measurements, which indicate that the gaps are isotropic, suggests that the symmetry is more likely to be extended-$s_\pm$ wave in character. 
\end{abstract}


\end{frontmatter}
\section{Introduction}
Since the discovery of superconductivity in iron arsenide compounds \cite{Kamihara:2008p7994,Takahashi:2008p7382,Ren:2008p8174}, neutron scattering experiments have made significant contributions to our understanding of the underlying physics.  Early neutron diffraction results generated considerable excitement because they revealed remarkable similarities with the high-temperature copper oxide superconductors.  For example, in both the iron arsenides and the cuprates, superconductivity arises when an antiferromagnetically ordered phase has been suppressed by chemical doping \cite{DeLaCruz:2008p8095}.  Neutron scattering continues to be essential in determining the magnetic and structural phase diagrams of these materials as a function of dopant concentration or applied pressure \cite{Lynn:2009p18084}.  On the other hand, neutrons have also identified important differences with the cuprates, such as the reduced size of the ordered moments and their extreme sensitivity to structural modifications \cite{DeLaCruz:2008p8095,Rotter:2008p8836,Yildirim:2008p11667}.  Elastic neutron scattering, which probes static magnetic and structural correlations, is discussed in more detail in another article in this issue \cite{Lynn:2009p18084}.  The purpose of this review is to summarize the results of inelastic neutron scattering, which probes dynamic correlations involving phonons and spin fluctuations, both of which are candidates for binding the superconducting electron pairs.

With the discovery of any new family of superconductors, the first task is to determine whether the critical temperature can be explained by electron-phonon coupling within a conventional BCS theory.  This question is usually addressed within the formalism of Eliashberg theory \cite{Carbotte:1990p7391}, in which the superconducting energy gap is expressed in terms of a spectral density  function derived from the phonon density-of-states (PDOS) weighted by electron-phonon matrix elements.  We will review inelastic neutron scattering, mostly on polycrystalline samples, that have been used to estimate the PDOS \cite{Christianson:2008p13379} and validate the results of first principles density functional calculations \cite{Singh:2008p8173,Boeri:2008p8844}.  In broad terms, the agreement between theory and experiment is very good, although some modes are extremely sensitive to the spin state assumed in the theoretical estimates.  Although there are subtle, so far unexplained, anomalies that will require further single crystal measurements to resolve fully, the early consensus is that the electron-phonon coupling is too weak by a factor of about five to explain the observed critical temperatures.

If phonons are not responsible for the superconductivity, spin fluctuations offer an alternative bosonic spectrum to mediate the electron pairing.  As first shown by de la Cruz \textit{et al} \cite{DeLaCruz:2008p8095}, the ground state of the non-superconducting parent compounds is a spin density wave, whose transition occurs close to a tetragonal-orthorhombic structural transition.  Both chemical doping and, in contrast to the cuprates, pressure \cite{Park:2008p8883} can be used to suppress antiferromagnetic order and induce superconductivity.  One of the key questions to resolve is what drives the magnetism.  Band structure calculations show that hole pockets at the $\Gamma$-point and the electron pockets at the $M$-point can show strong nesting with a sharp peak in the Lindhard susceptibility at Q=($\pi$,$\pi$), using tetragonal notation \cite{Singh:2009p15651}.  Since this is also the wavevector of magnetic order, it is natural to propose that the undoped arsenides are itinerant spin density waves like chromium, which would provide an explanation for the reduced size of the ordered moment.  However, some have argued that the pnictides are in fact close to a Mott insulating phase and that the reduced moments are due to frustration caused by competing superexchange interactions \cite{Si:2008p12538,Fang:2008p8200}, so the strength of electron correlations remains an important issue.  Inelastic neutron scattering experiments have shown that the spin waves are strongly dispersive, with velocities that are not inconsistent with an itinerant spin density wave, and more three-dimensional than the cuprates.  They also show substantial energy gaps that are not fully understood.

Finally, we review three reports of a spin resonant excitation seen so far only in the `122' compounds \cite{Chi:2008p14982,Christianson:2008p14965,Lumsden:2008p14911}.  With chemical doping, the spin fluctuations become more diffuse in the normal state, though still centered at the antiferromagnetic wavevectors.  However, below T$_c$, these fluctuations condense into a resonant excitation that is localized in both momentum transfer, $Q$, and energy transfer, $\omega$.  Such resonant excitations have been observed in a wide range of high-temperature copper oxide superconductors \cite{Hufner:2008p9494} as well as, more recently, several heavy fermion superconductors \cite{Metoki:1998p18348,Sato:2001p11225,Stock:2008p6655,Stockert:2008p11663}, where they are considered to be evidence of $d$-wave superconductivity \cite{Chang:2007p11651,Norman:2007p9025}.  In the case of the iron arsenides, a comparison with ARPES data suggests that the resonance is evidence of extended-$s_\pm$ wave symmetry \cite{Kuroki:2008p12581,Mazin:2008p11687}, in which the disconnected hole and electron pockets have energy gaps of opposite sign.  Since there is no angular anisotropy of the gap in this symmetry, it is difficult to verify by the techniques used in the copper oxides.  Inelastic neutron scattering is so far the only probe that has provided phase-sensitive information about the unconventional symmetry of the energy gap in the iron arsenide superconductors.

\section{Lattice Dynamics}
Within Eliashberg theory, the superconducting gap equation is expressed in terms of an electron-phonon coupling, $\lambda$, and a Coulomb repulsion, $\mu^\ast$ \cite{Carbotte:1990p7391}.  The electron-phonon coupling is derived from the spectral density function $\alpha^2 F(\omega)$, in which the phonon modes, represented by the bare phonon density-of-states (PDOS), $F(\omega)$, are weighted by electron-phonon matrix elements.  In many superconductors, there is little difference in the functional forms of $\alpha^2 F(\omega)$ and $F(\omega)$, \textit{i.e.}, the phonons are all coupled to the electronic states equally strongly, but there are exceptions.  For example, in MgB$_2$, the electron-phonon coupling of the $E_{2g}$ modes is particularly strong because their energies are governed by strongly covalent boron-boron $\sigma$-bonds.  In principle, $\alpha^2 F(\omega)$ can be determined by inverting tunneling data, but this is not always available so neutron scattering measurements of $F(\omega)$ play a valuable role in providing initial estimates of T$_c$ and in validating \textit{ab initio} calculations.

In BCS superconductors, numerical calculations based on the Eliashberg equations have shown that the critical temperature can often be approximated using the Allen-Dynes equation \cite{Allen:1975p16205}
\begin{equation} 
   \label{ADTc} 
      k_{B}T_{c} = \frac{\hbar \omega_{ln}}{1.2}
      \exp{\left[-\frac{1.04(1+\lambda)}{\lambda-\mu^*(1+0.62\lambda)}\right]}
\end{equation}
where $\omega_{ln}$ is a logarithmic phonon average defined in equation 2.16 of ref. \cite{Carbotte:1990p7391}, $\mu^\ast$ is the Coulomb repulsion, and the electron-phonon coupling, $\lambda$, is given by
\begin{equation} 
   \label{lambda} 
      \lambda = 2\int_0^\infty{d\omega\frac{\alpha^2 F(\omega)}{\omega}}
 \end{equation} 
 The Coulomb repulsion, $\mu^\ast$, is normally treated as a phenomenological parameter but it is possible to make accurate predictions of the critical temperature from first principles calculations without any adjustable parameters \cite{Floris:2005p16127}. 
 
There were two \textit{ab initio} calculations of the phonon density-of-states in LaFeAsO before the first neutron measurements were published, and both are in broad agreement with each other \cite{Boeri:2008p8844,Singh:2008p8173}.  Singh and Du established the main features of the band structure in the Local Spin Density Approximation, showing that the low-lying electronic states arise from five iron $d$-bands spanning an energy range of -2.1\,eV to 2.0\,eV, with the oxygen and arsenic $p$-states well below the Fermi level \cite{Singh:2008p8173}.  The Fermi surface consists of two-dimensional cylinders, with two hole pockets at the $\Gamma$-point and two electron pockets at the M-point, derived mainly from $d_{xz}$ and $d_{yz}$ states, along with a fifth more three-dimensional hole pocket, centered at Z, derived from $d_{z^2}$ states hybridized with As $p$-states and La orbitals.  

The resulting phonon excitations are spread over 70\,meV, with those over 40\,meV representing oxygen modes.  Those at lower frequency are of mixed lanthanum, iron, and arsenic character.   As several of the electron bands are two-dimensional, it is not surprising that the optic modes show little dispersion along the $\Gamma$-Z direction, but the acoustic phonons are fairly three-dimensional in character with a Debye temperature of 340\,K.  The calculated phonon density-of-states has three main peaks at approximately 12, 21, and 34\,meV.

On the basis of such band structure calculations, Boeri \textit{et al} calculate the electron-phonon coupling, which they find to be evenly distributed among all the modes \cite{Boeri:2008p8844}.  When they integrate this coupling over all frequencies, they derive $\lambda=0.21$, which is much smaller than any other known electron-phonon superconductors.  In the Allen-Dynes approximation, they estimate T$_c$ = 0.5\,K, from their calculated value of $\omega_{ln}$ = 205\,K, assuming $\mu^\ast=0$.   A more accurate numerical Migdal-Eliashberg calculation only increases this to 0.8\,K, and they estimate that $\lambda$ would have to be a factor five stronger to generate the observed T$_c$ of 26\,K.  Their calculations are based on the non-superconducting parent compound, LaFeAsO, but adding electrons through fluorine doping would, in the rigid band approximation, reduce the electronic density-of-states at the Fermi level, and so tend to reduce T$_c$ even further.  Orbitals that might be expected to produce an enhanced electron-phonon coupling, such as the $d_{x^2-y^2}$ orbitals that are directed along the Fe-Fe bond, are too far from the Fermi level to have a strong influence.

\begin{figure}[tb]
   \label{LaFeAsOF_ARCS}
   \centering
   \includegraphics[width=\columnwidth,clip]{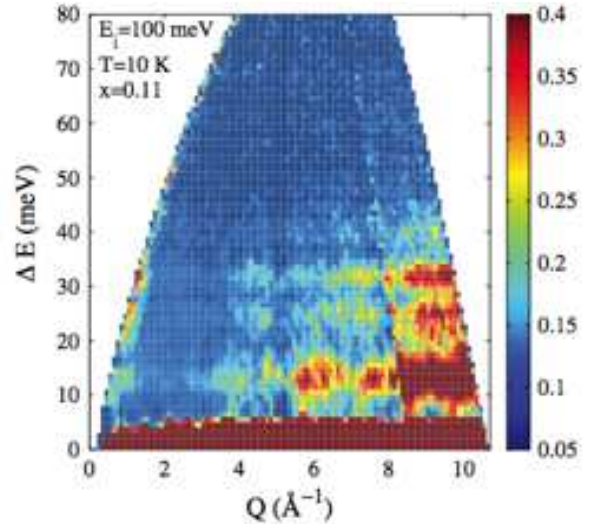}
   \caption{Inelastic neutron scattering from LaFeAsO$_{0.89}$F$_{0.11}$ as a function of momentum transfer, Q, and energy transfer, $\Delta$E (denoted as $\omega$  in the text), measured at 10\,K with an incident neutron energy of 100\,meV  \cite{Christianson:2008p13379}. The units of the intensity scale are arbitrary.}
\end{figure}

\begin{figure}[tb]
   \label{LaFeAsOF_PDOS}
   \centering
   \includegraphics[width=0.9\columnwidth,clip]{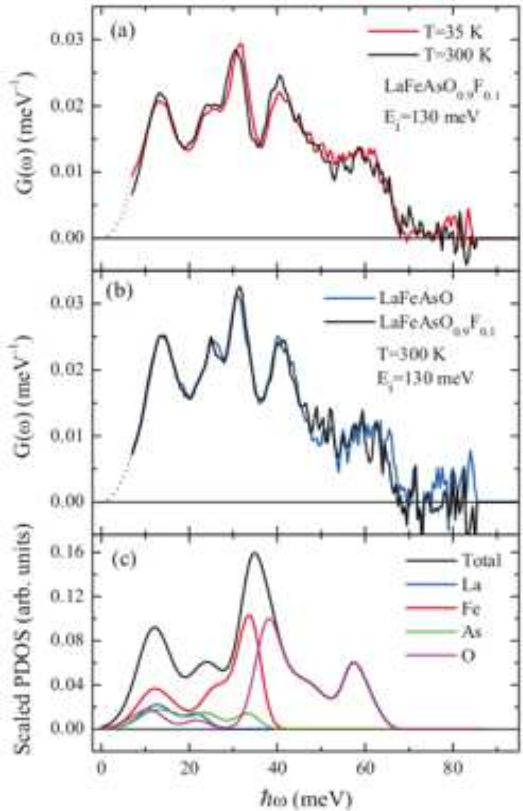}
   \caption{Generalized phonon density of states, $G(\omega)$, of LaFeAsO$_{1-x}$F$_{x}$ measured with an incident neutron energy of 130\,meV and normalized to an integral of 1.0 after correction for the Bose population and Debye-Waller factors, detector efficiency, and multiphonon scattering \cite{Christianson:2008p13379}. (a) Experimental $G(\omega)$ measured at 35 and 300\,K for $x=0.1$. (b) Experimental $G(\omega)$ for $x=0$ and $x=0.1$ measured at 300\,K. (c) First-principles calculation of the phonon density-of-states of LaFeAsO based on the band structure of Singh and Du \cite{Singh:2008p8173}.}
   \vspace{-0.2in}
\end{figure}

Even before any measurements were reported, it therefore seemed unlikely that a conventional BCS mechanism could explain the elevated transition temperatures in the iron arsenides.  Nevertheless, it was important to validate these predictions from experiment.  Although inelastic x-ray scattering can also be used to measure phonon dispersion relations (and has been used in the iron pnictides \cite{Fukuda:2008p13406,Reznik:2008p13555}), neutron scattering is an extremely efficient method of determining the phonon density-of-states.  In a multicomponent system, the inelastic neutron scattering law in polycrystalline samples is given by
\begin{equation}\begin{split}
   \label{Sqw} 
       {\rm S}(Q,\omega) = \sum_{i}
       &{\sigma_i\frac{\hbar Q^2}{2M_i}\exp{(-2W_i(Q))}}\\
       &\frac{{\rm G}_i(\omega)}{\omega}[n(\omega)+1]
\end{split}\end{equation} 
where $\sigma_i$ and $M_i$ are the neutron scattering cross section and 
atomic mass of the $i^{th}$ atom and $n(\omega)=[\exp{(\hbar\omega/k_{B}T)}-1]^{-1}$ 
is the Bose population factor.  The generalized PDOS, ${\rm G}(\omega) = \sum_i{{\rm G}_i(\omega)}$, where G$_i$($\omega$) is defined as 
\begin{equation}
   \label{GDOS} 
      {\rm G}_i(\omega) = \frac{1}{3N}\sum_{j{\bf q}}
      {|{\bf e}_i(j,{\bf q})|^2\delta[\omega-\omega(j,{\bf q})]} 
\end{equation}
which, in turn, defines the Debye-Waller factor, W$_i$(Q) through
\begin{equation}
   \label{DBW} 
      W_i(Q) = \frac{\hbar Q^2}{2M_i}
      \int_0^\infty{d\omega\frac{G_i(\omega)}{\omega}[2n(\omega)+1]}
\end{equation}
$\omega(j,{\bf q})$ and $e_i(j,{\bf q})$ are the frequencies and eigenvectors, respectively, of the phonon modes.

Inelastic neutron scattering therefore measures a sum of the partial phonon density-of-states of each constituent element weighted by $\sigma_i$/M$_i$.  In LaFeAsO, these weights, relative to the value for oxygen, are 0.23, 0.76, and 0.27, for La, Fe, and As, respectively \cite{Christianson:2008p13379}, so the low energy scattering is dominated by iron modes.  The generalized PDOS also differs from the base PDOS, F($\omega$), because of additional weighting by the eigenvectors (see Equation \ref{GDOS}).  Osborn \textit{et al} argued that this makes the use of G($\omega$) preferable to F($\omega$) as an approximation to $\alpha^2$F($\omega$), because these eigenvectors also enter into electron-phonon matrix elements  \cite{Osborn:2001p49}.  Nevertheless, the safest way to compare \textit{ab initio} calculations to the neutron data is to calculate the generalized PDOS directly as was done by Bohnen \textit{et al} in the case of MgB$_2$ \cite{Bohnen:2001p11933}.

Strictly speaking, Equation \ref{Sqw} only applies to materials in which the neutron cross sections of the constituent elements, $\sigma_i$, are entirely incoherent, whereas both iron and arsenic have strongly coherent cross sections.  This means that there will be strong deviations from the simple $Q^2$ dependence of the scattering intensity caused by variations in the eigenvectors within each Brillouin zone.  Nevertheless, if the data are taken on polycrystalline samples, so that the measured scattering is spherically averaged, and integrated over a sufficiently broad range of $Q$, they can be used to derive a good approximation to the generalized PDOS even when the $\sigma_i$ are coherent.  This is known as the incoherent approximation, which must be satisfied for the analysis to be reliable.  In MgB$_2$, for example, there were reports of a low-energy peak that were then shown to be an artifact of insufficient Q-averaging \cite{Osborn:2001p49}.  However, most of the reported PDOS measurements in the iron arsenides were taken on neutron spectrometers using relatively high incident energies and summed over a large range of momentum transfers, ensuring that the conditions for the incoherent approximation are met.

\begin{figure*}[tb]
   \label{Phonons_122}
   \centering
   \includegraphics[width=1.8\columnwidth,clip]{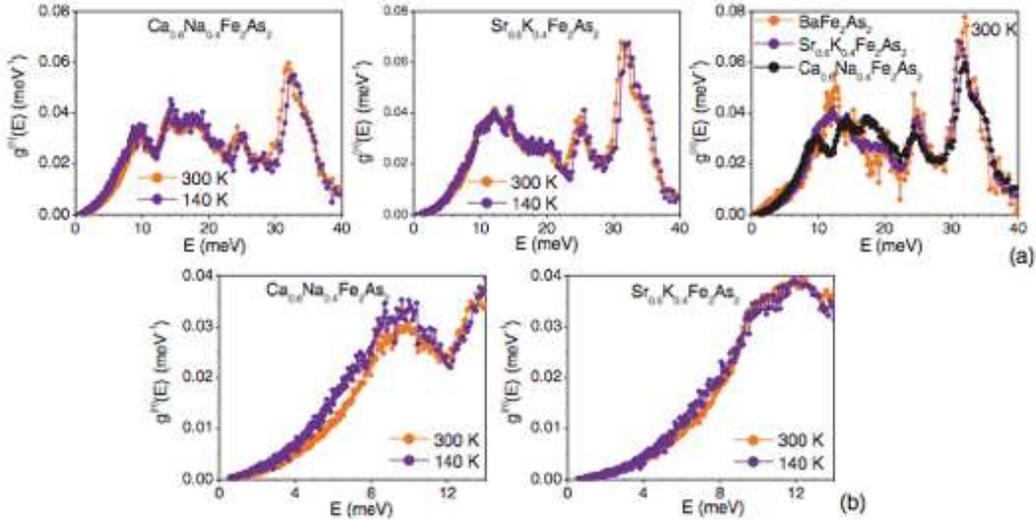}
   \caption{The experimental phonon spectra of Sr$_{0.6}$K$_{0.4}$Fe$_2$As$_2$ and Ca$_{0.6}$Na$_{0.4}$Fe$_2$As$_2$, measured in neutron energy gain with an incident neutron energy of 3.1\,meV \cite{Mittal:2008p15146}, compared to BaFe$_2$As$_2$, measured in energy loss with an incident energy of 57.5\,meV \cite{Mittal:2008p13172}. All the phonon spectra are normalized to unity.}
\end{figure*}

The first reported phonon measurements in the `1111' system covered a limited energy range up to 20\,meV \cite{Qiu:2008p12585}, but subsequent experiments have covered the entire phonon spectrum \cite{Christianson:2008p13379}.  Christianson \textit{et al} used samples of both non-superconducting LaFeAsO and superconducting LaFeAsO$_{1-x}$F$_x$, with $x\approx0.1$ prepared by two different synthesis groups, based at Oak Ridge National Laboratory and Ames Laboratory.  The data were collected on the newly commissioned Fermi chopper spectrometer, ARCS, at the Spallation Neutron Source in Oak Ridge.  Using incident neutron energies of 130, 60, and 30\,meV, supplemented by some triple-axis measurements, they observed peaks at 12, 25, 31, 40, and 60\,meV.  These features seem to be unaffected either by temperature reduction, apart from a sharpening of the peaks, or by dopant concentration.  In particular, there does not appear to be any strong phonon renormalization either at the orthorhombic transition in the non-superconducting samples, or at T$_c$ in the superconducting samples.

A comparison of their data with the \textit{ab initio} calculations showed good qualitative agreement in the overall energy scale of the phonons and the energies of most of the peaks.  The most notable discrepancy was in the location of the 31\,meV peak, which is distinctly softer than the theoretical prediction.  A similar discrepancy was noted in inelastic x-ray scattering, where it was attributed it to a 30\% reduction in the Fe-As force constants\cite{Fukuda:2008p13406}.  The length of the Fe-As bond shows a remarkable sensitivity to the iron spin state; calculations that underestimate the iron magnetism also substantially underestimate the bond lengths, and therefore overestimate the bond energies.  The fact that there is such little variation in the phonon peak energies is evidence therefore that, locally, the iron spin state is remarkably robust, persisting above the SDW transition temperature in LaFeAsO and surviving the destruction of SDW order in LaFeAsO$_{0.9}$F$_{0.1}$.  This is borne out by measurements of the spin dynamics discussed in the next section.

There have also been phonon measurements performed on the `122' compounds, firstly with experiments on non-superconducting BaFe$_2$As$_2$  \cite{Mittal:2008p13172} and then on superconducting Sr$_{0.6}$K$_{0.4}$Fe$_2$As$_2$ (T$_c$=32\,K) and Ca$_{0.6}$Na$_{0.4}$Fe$_2$As$_2$ (T$c$=21K) \cite{Mittal:2008p15146}.  The measurements were taken on the IN4 spectrometer, using an incident neutron energy of about 60\,meV, and the IN6 spectrometer, using an incident energy of 3.1\,meV, both at the Institut Laue Langevin, France.  The IN6 data were measured in neutron energy gain, which requires an elevated temperature.  With the absence of light oxygen ions in this structure, the maximum phonon energy is just under 40\,meV.  However, in other respects, the measured phonon spectra of the different `122' compounds look similar to the `1111' compounds and to each other.  There are peaks at 12, 22, 27, and 34\,meV in the barium and strontium samples but, in the calcium sample, the 22\,meV peak appears to have shifted down to below 18\,meV.  Since the bond lengths are shorter in Ca$_{0.6}$Na$_{0.4}$Fe$_2$As$_2$, which would normally increase the mode energies, there must be some change in the bonding characteristics.   Mittal \textit{et al} note that there is a slight stiffening from 300\,K to 140\,K of the higher energy peaks in the IN6 data, but a softening of the acoustic modes, which they suggest is a sign of electron-phonon coupling \cite{Mittal:2008p15146}.

In summary, neutron scattering studies of the lattice dynamics of the iron arsenides are broadly consistent with \textit{ab initio} calculations, although there are some unexplained anomalies that will require more detailed single crystal measurements before they are explained satisfactorily.  The integrated electron-phonon coupling is estimated to be far too weak to be responsible for the superconductivity.  However, the sensitivity of the Fe-As bond length to the iron spin state shows that there are potential sources of electron-phonon coupling that may need to be investigated more fully before some contribution from electron-phonon coupling is definitively ruled out.

\section{Spin Dynamics}
The non-superconducting parent compounds, such as LaFeAsO or BaFe$_2$As$_2$, undergo two phase transition with decreasing temperature \cite{DeLaCruz:2008p8095,Rotter:2008p8836,Huang:2008p14968}.  The first is a structural transition from the high-temperature tetragonal phase to a low-temperature orthorhombic phase, which is closely followed by (or sometimes coincident with) a second magnetic transition \cite{Lynn:2009p18084}.  The low-temperature antiferromagnetic structure is not the checkerboard order that would be expected from nearest-neighbor antiferromagnetic interactions, but a stripe phase, which implies the presence of competing interactions (see Fig. 4).  Yildirim has shown that this stripe structure results from an inherent frustration produced by a strong next-nearest-neighbor antiferromagnetic exchange between iron spins on the square planar lattice \cite{Yildirim:2008p11683}.  The observed structure is stable provided the nearest-neighbor exchange, $J_1$, along the square edges, and next-nearest-neighbor exchange, $J_2$, along the square diagonals, satisfy, $J_2>J_1 /2$.  He postulates that the structural transition is a means of relieving this frustration, which implies the existence of strong short-range spin correlations above the antiferromagnetic transition temperature, T$_{SDW}$, consistent with the discussion in the previous section.

\begin{figure}[tb]
   \label{Magnetic_Order}
   \centering
   \includegraphics[width=0.8\columnwidth,clip]{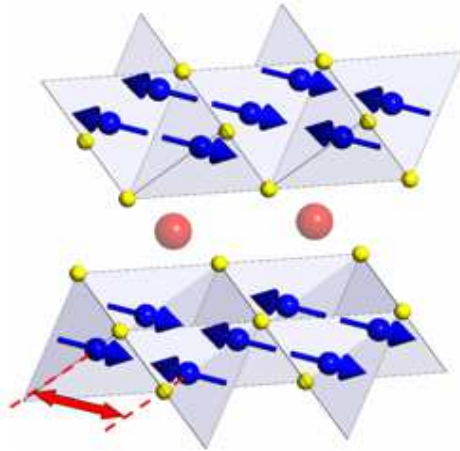}
   \caption{The crystal and magnetic structure of BaFe$_2$As$_2$ (taken from Ref. \cite{Christianson:2008p14965}). The unit cell contains two layers of Fe$_2$As$_2$ tetrahedra (Fe, blue spheres; As, yellow spheres), separated by planes of barium atoms (red spheres). The blue arrows show the observed ordering of the iron spins. The red arrow shows the spacing of the antiferromagnetic stripes.}
\end{figure}

As stated in the introduction, there are two alternative explanations for the origin of the magnetic interactions.  The first is that the antiferromagnetism is produced by a nesting instability coupling the $\Gamma$-centered hole pockets and the M-centered electron pockets.  This assumes an itinerant picture of weakly interacting electrons, that is consistent with the reduced size of the magnetic moments (0.36$\mu_B$ in LaFeAsO \cite{DeLaCruz:2008p8095} or 0.87$\mu_B$ in BaFe$_2$As$_2$ \cite{Huang:2008p14968}), which can be explained by density functional theory provided the experimental lattice parameters are used \cite{Mazin:2008p12399}.  However, there are alternative models, which assume much stronger electron correlations, with moments reduced by frustration \cite{Si:2008p12538,Fang:2008p8200}.  

Inelastic neutron scattering is the most direct way of determining the spin wave excitations of the antiferromagnetically ordered compounds.   Whatever the origin of the magnetic interactions, it is usually possible to analyze the measured dispersion using a Heisenberg model including single-ion anisotropy terms that are required to produce the observed energy gaps \cite{Ewings:2008p15404,Yao:2008p7354}

\begin{equation}\begin{split}
   \label{Heisenberg} 
      H = &\sum_{ij}{J_{ij}\mathbf{S_i}\cdot\mathbf{S_j}} + \\
      &\sum_i{K_c (S^z_i)^2 + K_{ab}\left[(S^x_i)^2-(S^y_i)^2\right]}
\end{split}\end{equation}

Ewings \textit{et al} derive explicit solutions of this Hamiltonian,  giving both the energies and spin wave cross sections \cite{Ewings:2008p15404}, but, at low energies,  the spin wave dispersion can be approximated by \cite{McQueeney:2008p15157} 

\begin{equation}\begin{split}
   \label{Dispersion} 
      \hbar\omega(\mathbf{q}) = \sqrt{\Delta^2+v^2_{xy}(q^2_x+q^2_y) + v^2_z q^2_z}
\end{split}\end{equation}
where \textbf{q} is the reduced wavevector relative to the antiferromagnetic zone center, $\Delta$ is the anisotropy gap, and $v_{xy}$ and $v_z$ are the in-plane and $c$-axis spin wave velocities.

\begin{figure}[tb]
   \label{BaFe2As2_MERLIN}
   \centering
   \includegraphics[width=\columnwidth,clip]{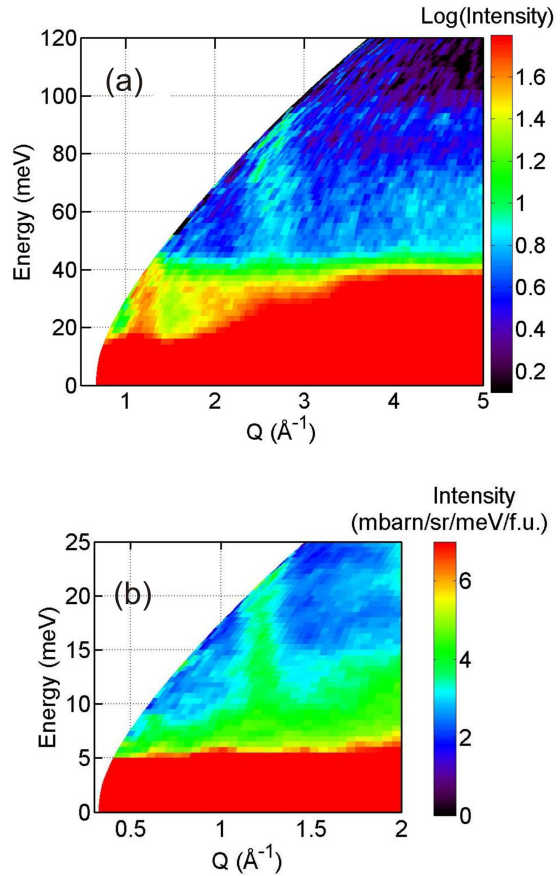}
   \caption{Neutron scattering spectra from a polycrystalline sample of BaFe$_2$As$_2$ at 7\,K \cite{Ewings:2008p15404}.  The data were taken on the MERLIN (ISIS) spectrometer using incident neutron energies of (a) 200\,meV and (b) 50\,meV.  The pillars of scattering show the steep spin wave excitations emerging from the ($\frac{1}{2}$,$\frac{1}{2}$,$l$) and ($\frac{1}{2}$,1,$l$) positions (using the tetragonal Brillouin zone) at Q=1.2\,\AA\ and 2.6\,\AA, respectively.}
\end{figure}

\begin{figure}[tb]
   \label{SrFe2As2_spinwaves}
   \centering
   \includegraphics[width=\columnwidth,clip]{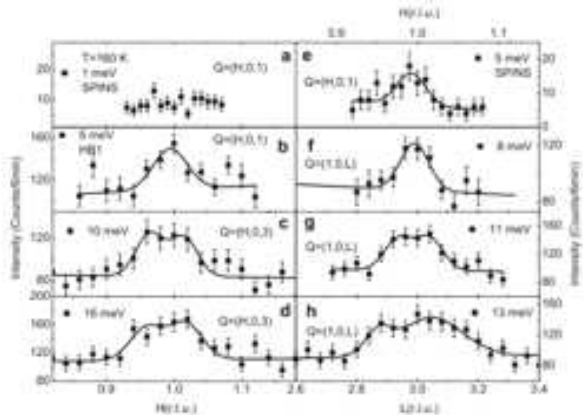}
   \caption{Constant energy scans performed on triple-axis spectrometers on SrFe$_2$As$_2$ at 160\,K, showing the broadening as a function of increasing energy resulting from the spin wave dispersion \cite{Zhao:2008p13418}.}
\end{figure}

Spin wave measurements have been reported in a number of the `122' compounds, SrFe$_2$As$_2$ \cite{Zhao:2008p13418}, CaFe$_2$As$_2$ \cite{McQueeney:2008p15157,Diallo:2009p15798}, and BaFe$_2$As$_2$ \cite{Ewings:2008p15404,Matan:2008p13460}.  Most of these are single crystal measurements using triple-axis spectrometers, although Ewings \textit{et al} \cite{Ewings:2008p15404} and Diallo \textit{et al} \cite{Diallo:2009p15798} used pulsed neutron source Fermi chopper spectrometers, MERLIN and MAPS respectively, at ISIS, to extend the energy range to 100\,meV and above.  Broadly speaking, all these experiments are consistent.  They observe extremely steep spin waves emerging from the SDW wavevector with a substantial energy gap, ranging from 6.9\,meV in SrFe$_2$As$_2$ \cite{Zhao:2008p13418} to 9.8\,meV in BaFe$_2$As$_2$ \cite{Matan:2008p13460}.  The value of the spin wave velocity is more uncertain because it is difficult to resolve the propagating modes  in constant energy scans when the dispersions are so steep, except at very high energy \cite{Diallo:2009p15798}.  However, estimates can be derived by performing model simulations including the instrumental resolution.  The in-plane estimates vary from 280\,meV\AA\  in BaFe$_2$As$_2$ \cite{Ewings:2008p15404,Matan:2008p13460} to 420\,meV\AA\  in CaFe$_2$As$_2$ \cite{McQueeney:2008p15157}.  Low-energy measurements predicted zone boundary  energies of about 175\,meV \cite{Ewings:2008p15404,Matan:2008p13460}, which is consistent with the first single crystal measurements using pulsed neutrons  \cite{Diallo:2009p15798}, although the spin waves above 100\,meV appear to be heavily damped.  The out-of-plane velocities are smaller but still substantial, reduced from the in-plane values by a factor of 2 \cite{McQueeney:2008p15157} to 5 \cite{Matan:2008p13460}, so the magnetic order is truly three-dimensional.  The temperature dependence of the spin wave scattering is consistent with Bose statistics, with the energy gap renormalizing to zero at T$_{SDW}$.  However, there is evidence of the persistence of two-dimensional short-range spin correlations in BaFe$_2$As$_2$ above T$_{SDW}$, in the form of quasielastic scattering centered on ($\frac{1}{2}$,$\frac{1}{2}$,$l$) rods \cite{Matan:2008p13460}.  

If the Heisenberg model includes next-nearest-neighbors, as required to stabilize the stripe structure, there are four exchange constants: $J_{1a}$ and $J_{1b}$ couple antiferromagnetic and ferromagnetic nearest neighbors, respectively; $J_2$ couples next-nearest neighbors within the plane; and $J_z$ is the out-of-plane coupling.  In SrFe$_2$As$_2$, Zhao \textit{et al} estimated the exchange interactions to be $J_{1a}+2J_2\sim100$\,meV, with $J_z\sim5$\,meV \cite{Zhao:2008p13418}.  These values are consistent with McQueeney \textit{et al}'s analysis of CaFe$_2$As$_2$, where they estimate $J_{1a}$ = 41\,meV, $J_{1b}$ = 10\,meV, $J_2$ = 21\,meV, and $J_z$ = 3\,meV.  Note that the value of $J_2$ is sufficiently large to stabilize stripe antiferromagnetism.

In an itinerant SDW, the spin wave velocity produced by the excitation of particle-hole pairs across the nested surfaces is given by $\sqrt{v_e v_h}$, where $v_e$ and $v_h$ are the electron and hole velocities of their respective bands \cite{Fawcett:1988p15790}.  In LaFeAsO, band structure calculations estimate that the in-plane electron velocity is a factor of 3 greater than the in-plane hole velocity, which is, in turn, a factor of 2.5 greater than both the $c$-axis hole and electron velocities \cite{Singh:2008p8173}. The absolute values are uncertain since the calculated bandwidths are greater than ARPES measurements \cite{Liu:2008p14078}.  They do, however, give a ratio of in-plane to $c$-axis spin wave velocities of about 4, which is within the experimental range.  Zhao \textit{et al} go further and show that the absolute values are in reasonable agreement with band structure, after renormalizing the bandwidths \cite{Zhao:2008p13418}.  

The damping of the high-energy excitations seen in CaFe$_2$As$_2$ could be due to the decay of spin waves into electron-hole pairs within a Stoner continuum  \cite{Diallo:2009p15798}.  More speculatively, Matan \textit{et al} report the existence of excess scattering above 20\,meV, which is more two-dimensional than the spin waves, and which they suggest is consistent with the existence of excess scattering in the Stoner continuum.  However, the data from Ewings \textit{et al} do not show evidence for this anomaly \cite{Ewings:2008p15404}, so it is difficult to draw any firm conclusions.  Finally, there are claims that an unusual temperature dependence of the excitations in LaFeAsO is also consistent with itinerant spin density waves \cite{Ishikado:2008p13337}.

In summary, the neutron measurements in the parent compounds of the iron arsenides provide a consistent picture of spin wave dispersions, that are extremely steep and anisotropic, although clearly three-dimensional, with a very large energy gap.  The exchange parameters are consistent with the observed stripe phase, and are not inconsistent with itinerant SDW models, although it is too early to rule out more localized magnetic models based on strong but frustrated superexchange interactions.

\section{Resonant Spin Excitations}

\begin{figure*}[t]
   \label{Resonance_MERLIN}
   \centering
   \includegraphics[width=2\columnwidth,clip]{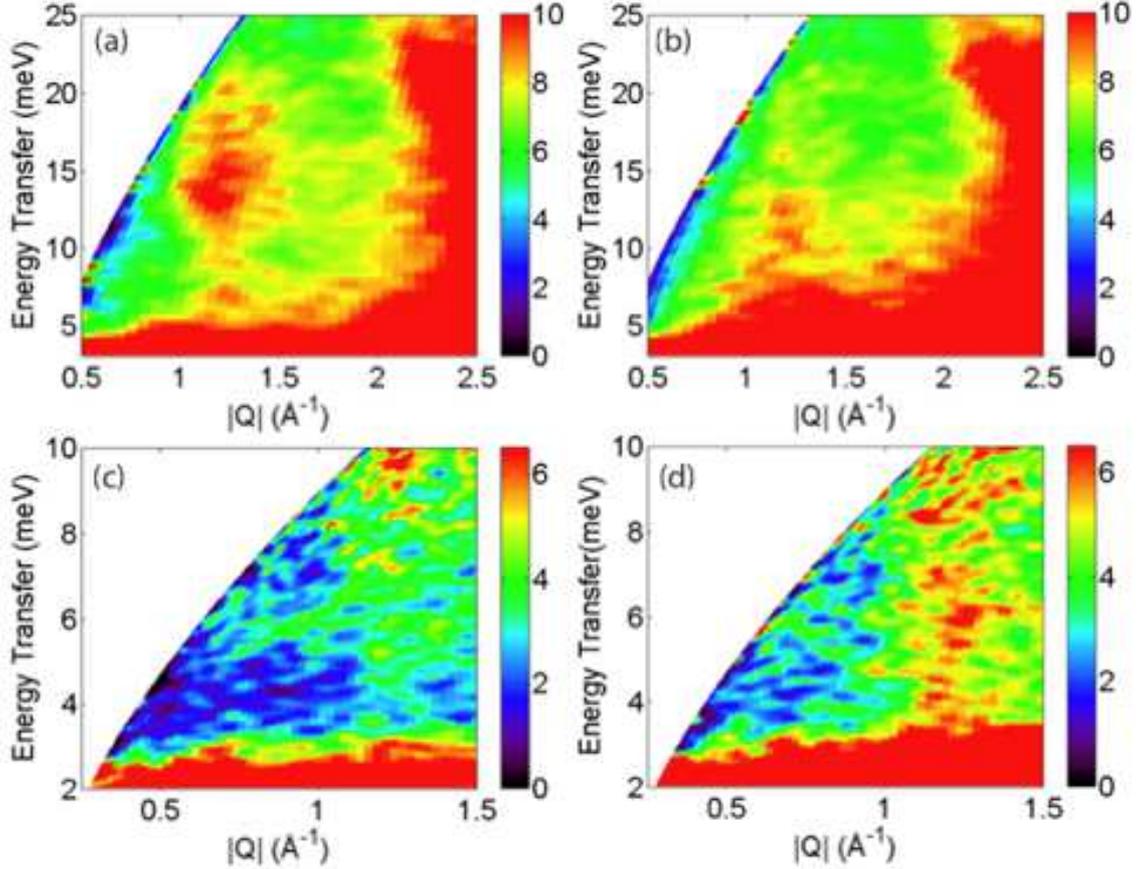}
   \caption{Inelastic neutron scattering on  Ba$_{0.6}$K$_{0.4}$Fe$_2$As$_2$, measured using an incident neutron energy of 60\,meV below and above T$_c$, \textit{i.e.},  7\,K (a) and 50\,K (b) respectively, and 15\,meV below and above T$_c$, (c) and (d) respectively.  The data shows the transfer of spectral weight from diffuse spin fluctuations centered at 1.15\,\AA$^{-1}$ in the normal state into a resonant spin excitation at an energy transfer of 15\,meV in the superconducting state.  The color scale is in units of mbarns/sr/meV/mol \cite{Christianson:2008p14965}.}
\end{figure*}

The previous two sections concerned inelastic neutron scattering results that provide the context within which superconductivity develops in the iron arsenides, but do not directly address the superconducting state itself.  That has been the traditional role for neutron scattering since, in conventional superconductors, neutrons do not couple directly to the superconducting order parameter.  However, neutron scattering results on other unconventional superconductors over the past twenty years has shown that they can provide direct information concerning the superconducting energy gap, in particular, its symmetry.  Surprisingly, in the case of the iron arsenides, it is currently the only technique to give phase-sensitive evidence of the gap symmetry, since many of the techniques commonly used in other superconductors do not apply.

Inelastic neutron scattering measures directly the dynamic magnetic susceptibility of the band electrons.  Normally, this signal is too diffuse for useful measurements, but there are cases where the magnetic response is strongly enhanced at particular energies and wavevectors making such experiments feasible.  The most common example is in transition metal magnets such as iron and cobalt, where the RPA susceptibility of the itinerant $d$-electrons has poles corresponding to propagating spin wave modes \cite{Cooke:1980p18413}.  In an itinerant SDW model, the spin waves described in the previous section fall into this category.  No such divergences exist in the dynamic magnetic susceptibility of conventional $s$-wave superconductors, but it turns out that they can exist when the superconducting energy gap changes sign, either within a single Fermi surface or between two disconnected Fermi surfaces.

Specifically, there is a strong enhancement of the dynamic magnetic susceptibility due to coherence factors that appear because of the anomalous Green function of the superconducting phase.  The following term appears in the non-interacting susceptibility \cite{Schrieffer:1964p18081,Fong:1995p18080,Norman:2007p9025,Chang:2007p11651}.  

\begin{equation}
   \label{Coherence} 
      \Big(1-\frac{\xi_{\mathbf{k}+\mathbf{q}}\xi_{\mathbf{k}}+\Delta_{\mathbf{k}}\Delta_{\mathbf{k}+\mathbf{q}}}{E_{\mathbf{k}+\mathbf{q}}E_{\mathbf{k}}}\Big)
\end{equation}
where the energy of the superconducting quasiparticles, $E_{\mathbf{k}} = \sqrt{\xi_{\mathbf{k}}^2+\Delta_{\mathbf{k}}^2}$.  Here, $\xi_\mathbf{k}$ are the single electron energies of the normal-state and $\Delta_\mathbf{k}$ are the values of the energy gap at points $\mathbf{k}$ on the Fermi surface.  

Equation \ref{Coherence} defines the principal characteristics of the resonant spin excitations.  If $\Delta_{\mathbf{k}+\mathbf{Q}} = \Delta_{\mathbf{k}}$, as in a conventional $s$-wave superconductor, Equation 8 vanishes on the Fermi surface ($\xi_{\mathbf{k}}=0$).  In this case, there is no pole when this non-interacting susceptibility is introduced into an RPA expression for the interacting susceptibility. However, if $\Delta_{\mathbf{k}+\mathbf{Q}} = -\Delta_{\mathbf{k}}$, \textit{i.e.} when \textbf{Q} connects two Fermi surface points whose superconducting order parameters have opposite sign, then Equation 8 is maximal, resulting in a pole at \textbf{Q} for some energy less than $2\Delta$.

Such resonant excitations are now believed to be a universal feature of the high-temperature copper oxide superconductors \cite{Hufner:2008p9494}, with observations in a large number of different systems \cite{RossatMignod:1991p10509,Mook:1993p10706,Fong:1995p18080,Fong:1999p11512,Dai:2000p10619,He:2002p11430}.   The resonance energy scales approximately as $\omega_0\sim5k_B$T$_c$.  They are commonly taken as evidence of $d_{x^2-y^2}$ symmetry, in which the energy gap changes sign within a single Fermi surface.  Strikingly, there are now several reports that similar resonant excitations are present in heavy fermion superconductors \cite{Metoki:1998p18348,Sato:2001p11225,Stock:2008p6655,Stockert:2008p11663}, where T$_c$ can be as low as 0.7\,K.  Nevertheless, the energy of the resonance appears to obey the same scaling relations, with $\omega_0/2\Delta$, where $\Delta$ is the maximum value of the superconducting energy gap, taking values between 0.62 and 0.74, a remarkable observation given that T$_c$ varies by over two orders of magnitude.

\begin{figure}[tb]
   \label{Resonance_Tdep}
   \centering
   \includegraphics[width=0.8\columnwidth,clip]{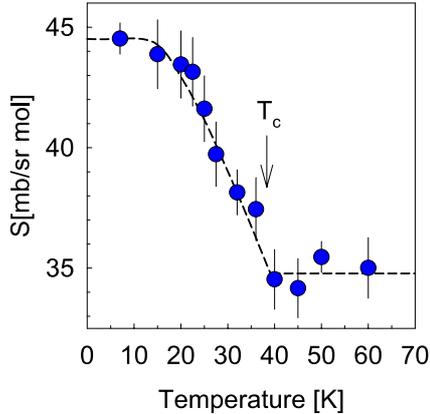}
   \caption{The inelastic neutron scattering from Ba$_{0.6}$K$_{0.4}$Fe$_2$As$_2$ integrated over a $Q$-range of 1.0 to 1.3\,\AA$^{-1}$ and an $\omega$ range of 12.5 to 17.5\,meV. The integration range corresponds to the region of maximum intensity of the resonant excitation observed below T$_c$ (see Fig. 7).  The dashed line is a guide to the eye below T$_c$ and shows the average value of the integrals above T$_c$ \cite{Christianson:2008p14965}.}
\end{figure}

\begin{figure}[tb]
   \label{resonance_theory}
   \centering
   \includegraphics[width=\columnwidth,clip]{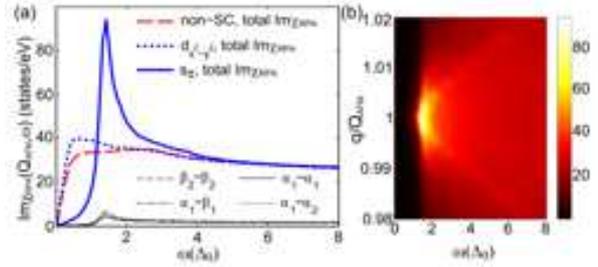}
   \caption{(a) Calculated imaginary part of the RPA spin susceptibility at the SDW wave vector $Q_{AFM}$ as a function of frequency in the normal and superconducting states. The red, dotted blue, and solid blue curves correspond to the total RPA susceptibility for non-superconducting, $d_{x^2-y^2}$ symmetry and extended-$s_\pm$ symmetry models.. The thin (black) curves refer to the partial RPA contributions for the interband and intraband transitions in the $s_\pm$ superconducting state. (b) Calculated imaginary part of the
total RPA spin susceptibility in the $s_\pm$ state as a function of frequency and momentum along \textbf{Q}=($h$,$h$) \cite{Korshunov:2008p13468}.}
\end{figure}

Figure 7 shows inelastic neutron scattering data taken on the MERLIN spectrometer, at the ISIS Pulsed Neutron Source.  The polycrystalline sample of  Ba$_{0.6}$K$_{0.4}$Fe$_2$As$_2$ used in these measurements was optimally doped with a T$_c$ of 38\,K.  The intense scattering at high Q and low energy results from phonon and elastic nuclear scattering, respectively, but the remaining scattering results from spin fluctuations.  In the normal phase, there is an echo of the sharp spin waves seen in Fig. 5, with a pillar of scattering at the antiferromagnetic wavevector, Q=1.15\AA\ .  However, the scattering is considerably more diffuse representing short-range spin fluctuations rather than propagating spin waves.      It also persists down to lower energy.  As the temperature is lowered through T$_c$, there is a transfer of spectral weight into the resonant spin excitation at an energy transfer of 15\,meV.  Since the maximum gap seen in ARPES data is 12\,meV \cite{Ding:2008p8816}, $\omega_0/2\Delta\sim0.58$ in good agreement with the scaling in other unconventional superconductors.  The temperature dependence of this resonant spin excitation is shown in Fig. 8, where it behaves like an order parameter, as also observed in the other unconventional superconductors.

The observation of a spin resonance does not necessarily imply $d$-wave symmetry.  In fact, ARPES data on Ba$_{0.6}$K$_{0.4}$Fe$_2$As$_2$ shows purely isotropic gaps around each surface \cite{Ding:2008p8816}, which is inconsistent with a $d$-wave model.  Although such models have been discussed in connection with the iron arsenides \cite{Kuroki:2008p12581}, most interest has been focussed on extended-$s_\pm$ models.  In fact, explicit calculations of the neutron scattering cross section predict the existence of a resonant spin excitation with extended-$s_\pm$, but not $d$-wave symmetry \cite{Korshunov:2008p13468,Maier:2008p11868}.  The small Fermi surfaces seen in the iron arsenides would not intersect any of the nodal lines in these models, so the gaps on each surface would be nearly constant.  However, the prediction is that the sign of the energy gap on the hole pockets at the $\Gamma$-point and the electron pockets at the M-point would be opposite, reflecting an electron-electron repulsion at short-range but an attractive interaction for electrons on neighboring iron atoms.  The wavevector connecting the hole and electron pockets is precisely where the resonant spin excitation has been observed.

More recently, resonant spin excitations have also been seen in single crystals of BaFe$_{1.84}$Co$_{0.16}$As$_2$ using both pulsed source and triple-axis instruments at Oak Ridge \cite{Lumsden:2008p14911}.  This has confirmed that the excitation is centered at Q=($\frac{1}{2}$,$\frac{1}{2}$) within the plane, but it is two-dimensional, with intensity spread out along Q=$l$ following a single-ion Fe$^{2+}$ form-factor.  Since T$_c$ = 22\,K, which is somewhat lower than in Ba$_{0.6}$K$_{0.4}$Fe$_2$As$_2$, the energy of the resonance is also lower at 9.6\,meV, in agreement with the previous scaling.  

There is a third report of a spin resonance in BaFe$_{1.9}$Ni$_{0.1}$A$_2$ (T$_c$ = 20\,K) \cite{Chi:2008p14982}, again observed at Q=($\frac{1}{2}$,$\frac{1}{2}$,$l$).  Chi \textit{et al} observe an energy dispersion of about 2\,meV; the resonance is at 9.1\,meV at $l = 0$ but at 7.0\,meV at $l = \pm1$.  They attribute this to antiferromagnetic coupling between the FeAs layers, although it could also reflect some three-dimensional modulation of the Fermi surfaces.  Although this work shows evidence of three-dimensionality in the superconducting order, it is still much more two-dimensional than the spin wave excitations of the parent compounds.

\section{Conclusions}
It is far too soon since the discovery of these fascinating compounds to declare that any of the important issues surrounding their superconductivity have been settled.  As shown in this review, there are many unanswered questions concerning the origin of the magnetic interactions, the strength of the electron-phonon coupling, or the dimensionality of the superconducting order parameter.  Nevertheless, it is remarkable how much progress has been made in such a short time.  This reflects the experience that the scientific community has gained over the past thirty years of studying unconventional superconductors, first the heavy fermions and then the copper oxide superconductors.  Many of the insights gained in those investigations have been directly applied to the new iron arsenide superconductors, and, in some cases, theories considered but then rejected in other systems have become useful here.  

This is particularly true for neutron scattering.  It took several years for the significance of the resonant spin excitations in the copper oxides to be appreciated.  Now our theoretical understanding is well advanced, particularly now that they have been seen in other unconventional superconductors.  This has allowed the new measurements on the iron arsenides to be incorporated into theories of the superconductivity much more rapidly than before.   Although the new methods of detecting extended $s_\pm$-wave symmetry have been proposed \cite{Inotani:2009p15561}, inelastic neutron scattering is until now the only phase-sensitive technique to provide direct evidence in these compounds.

In spite of the many similarities to other superconductors, the iron arsenides are not carbon copies.  If, as seems likely, the superconducting pairing is mediated by spin fluctuations, they are much more itinerant than in the copper oxides or the heavy fermions and the superconducting symmetry appears to be quite different.  The existence of the resonance at the same wavevector as the SDW does lend some support to the idea that the transition from antiferromagnetism to superconductivity occurs when the changes in the Fermi surface with doping or pressure suppress the nesting instability.  Although the conditions for magnetic order are no longer satisfied, the susceptibility is still sufficiently enhanced to favor spin fluctuation-mediated superconductivity as discussed by Singh in his review \cite{Singh:2009p15651}.  The fact that  the wavevectors characterizing both the spin density wave and superconductivity are the same is therefore no coincidence.

Inelastic neutron scattering experiments are only just beginning in these systems.  The improvements in sample quality and, particularly important for neutrons, size will allow more detailed studies over a wider range of wavevectors and energies, with single crystal results superceding the earlier polycrystalline data, as has already begun to happen. We have also not addressed the many materials science issues resulting from phase separation, both intrinsic and extrinsic.  Nevertheless, progress is likely to be rapid if the experience of the past year is an accurate guide.  

\section{Acknowledgements}
We acknowledge valuable conversations with Michael Norman, David Singh, and Taner Yildirim in writing this review, and express gratitude for our many collaborators over the past year. This work was supported by the Division of Materials Sciences and Engineering Division and the Scientific User Facilities Division of the Office of Basic Energy Sciences, U.S. Department of Energy Office of Science, under Contract Nos. DE-AC02-06CH11357 and DE-AC05-00OR22725.

%\bibliography{FeAs_INS}

\begin{thebibliography}{10}
\expandafter\ifx\csname url\endcsname\relax
  \def\url#1{\texttt{#1}}\fi
\expandafter\ifx\csname urlprefix\endcsname\relax\def\urlprefix{URL }\fi
\expandafter\ifx\csname href\endcsname\relax
  \def\href#1#2{#2} \def\path#1{#1}\fi

\bibitem{Kamihara:2008p7994}
Y.~Kamihara, T.~Watanabe, M.~Hirano, H.~Hosono, J Am Chem Soc 130 (2008)
  3296--3297.

\bibitem{Takahashi:2008p7382}
H.~Takahashi, K.~Igawa, K.~Arii, Y.~Kamihara, M.~Hirano, H.~Hosono, Nature 453
  (2008) 376--378.

\bibitem{Ren:2008p8174}
Z.-A. Ren, G.-C. Che, X.-L. Dong, J.~Yang, W.~Lu, W.~Yi, X.-L. Shen, Z.-C. Li,
  L.-L. Sun, F.~Zhou, Z.-X. Zhao, Europhys Lett 83 (2008) 17002.

\bibitem{DeLaCruz:2008p8095}
C.~D.~L. Cruz, Q.~Huang, J.~Lynn, J.~Li, W.~R. Ii, J.~Zarestky, H.~A. Mook,
  G.~F. Chen, J.~L. Luo, N.~Wang, P.~Dai, Nature 453 (2008) 899--902.

\bibitem{Lynn:2009p18084}
J.~W. Lynn, P.~Dai, arXiv:0902.0091.

\bibitem{Rotter:2008p8836}
M.~Rotter, M.~Tegel, D.~Johrendt, I.~Schellenberg, W.~Hermes, R.~P{\"o}ttgen,
  Phys Rev B 78 (2008) 020503.

\bibitem{Yildirim:2008p11667}
T.~Yildirim, arXiv:0807.3936.

\bibitem{Carbotte:1990p7391}
J.~P. Carbotte, Rev Mod Phys 62 (1990) 1027.

\bibitem{Christianson:2008p13379}
A.~D. Christianson, M.~D. Lumsden, O.~Delaire, M.~B. Stone, D.~L. Abernathy, M.~A.
  Mcguire, A.~S. Sefat, R.~Jin, B.~C. Sales, D.~Mandrus, E.~D. Mun, P.~C.
  Canfield, J.~Y.~Y. Lin, M.~Lucas, M.~Kresch, J.~B. Keith, B.~Fultz,
  E.~Goremychkin, R.~McQueeney, Phys Rev Lett 101 (2008) 157004.

\bibitem{Singh:2008p8173}
D.~Singh, M.-H. Du, Phys Rev Lett 100 (2008) 237003.

\bibitem{Boeri:2008p8844}
L.~Boeri, O.~V. Dolgov, A.~A. Golubov, Phys Rev Lett 101 (2008) 026403.

\bibitem{Park:2008p8883}
T.~Park, E.~Park, H.~Lee, T.~Klimczuk, E.~D. Bauer, F.~Ronning, J.~Thompson, J
  Phys-Cond Mat 20 (2008) 322204.

\bibitem{Singh:2009p15651}
D.~Singh, arXiv:0901.2149.

\bibitem{Si:2008p12538}
Q.~Si, E.~Abrahams, Phys Rev Lett 101 (2008) 076401.

\bibitem{Fang:2008p8200}
C.~Fang, H.~Yao, W.-F. Tsai, J.~Hu, S.~Kivelson, Phys Rev B 77 (2008) 224509.

\bibitem{Chi:2008p14982}
S.~Chi, A.~Schneidewind, J.~Zhao, L.~W. Harriger, Y.~Luo, G.~Cao,
  M.~Loewenhaupt, P.~Dai, arXiv:0812.1354.

\bibitem{Christianson:2008p14965}
A.~Christianson, E.~Goremychkin, R.~Osborn, S.~Rosenkranz, M.~D. Lumsden,
  C.~Malliakas, lllya Todorov, H.~Claus, D.~Y. Chung, M.~Kanatzidis, R.~Bewley,
  T.~Guidi, Nature 456 (2008) 930--932.

\bibitem{Lumsden:2008p14911}
M.~D. Lumsden, A.~Christianson, D.~Parshall, M.~B. Stone, S.~E. Nagler, H.~A.
  Mook, K.~Lokshin, T.~Egami, D.~L. Abernathy, E.~Goremychkin, R.~Osborn, M.~A.
  Mcguire, A.~S. Sefat, R.~Jin, B.~C. Sales, D.~Mandrus, arXiv:0811.4755.

\bibitem{Hufner:2008p9494}
S.~H{\"u}fner, M.~A. Hossain, A.~Damascelli, G.~Sawatzky, Rep Prog Phys 71
  (2008) 2501.

\bibitem{Metoki:1998p18348}
N.~Metoki, Y.~Haga, Y.~Koike, Y.~Oønuki, Phys Rev Lett 80 (1998) 5417--5420.

\bibitem{Sato:2001p11225}
N.~K. Sato, N.~Aso, K.~Miyake, R.~Shiina, P.~Thalmeier, G.~Varelogiannis,
  C.~Geibel, F.~Steglich, P.~Fulde, T.~Komatsubara, Nature 410 (2001) 340--343.

\bibitem{Stock:2008p6655}
C.~Stock, C.~Broholm, J.~Hudis, H.~J. Kang, C.~Petrovic, Phys Rev Lett 100
  (2008) 087001.

\bibitem{Stockert:2008p11663}
O.~Stockert, J.~Arndt, A.~Schneidewind, H.~Schneider, H.~S. Jeevan, C.~Geibel,
  F.~Steglich, M.~Loewenhaupt, Physica B 403 (2008) 973.

\bibitem{Chang:2007p11651}
J.~Chang, I.~Eremin, P.~Thalmeier, P.~Fulde, Phys Rev B 75 (2007) 24503.

\bibitem{Norman:2007p9025}
M.~Norman, Phys Rev B 75 (2007) 184514.

\bibitem{Kuroki:2008p12581}
K.~Kuroki, S.~Onari, R.~Arita, H.~Usui, Y.~Tanaka, H.~Kontani, H.~Aoki, Phys
  Rev Lett 101 (2008) 087004.

\bibitem{Mazin:2008p11687}
I.~Mazin, D.~Singh, M.~D. Johannes, M.~H. Du, Phys Rev Lett 101 (2008) 057003.

\bibitem{Allen:1975p16205}
P.~B. Allen, R.~C. Dynes, Phys. Rev. B 12 (1975) 905--922.

\bibitem{Floris:2005p16127}
A.~Floris, G.~Profeta, N.~Lathiotakis, M.~Luders, M.~Marques, C.~Franchini,
  E.~Gross, A.~Continenza, S.~Massidda, Phys Rev Lett 94 (2005) 037004.

\bibitem{Fukuda:2008p13406}
T.~Fukuda, A.~Baron, S.-I. Shamoto, M.~Ishikado, H.~Nakamura, M.~Machida,
  H.~Uchiyama, S.~Tsutsui, A.~Iyo, H.~Kito, J.~Mizuki, M.~Arai, H.~Eisaki,
  H.~Hosono, J Phys Soc Jpn 77 (2008) 103715.

\bibitem{Reznik:2008p13555}
D.~Reznik, K.~Lokshin, D.~C. Mitchell, D.~Parshall, W.~Dmowski, D.~Lamago,
  R.~Heid, K.~P. Bohnen, A.~S. Sefat, M.~A. McGuire, B.~C. Sales, D.~G.
  Mandrus, A.~Asubedi, D.~Singh, A.~Alatas, M.~H. Upton, A.~H. Said,
  Y.~Shvyd'ko, T.~Egami, arXiv:0810.4941.

\bibitem{Osborn:2001p49}
R.~Osborn, E.~Goremychkin, A.~Kolesnikov, D.~G. Hinks, Phys Rev Lett 87 (2001)
  017005.

\bibitem{Bohnen:2001p11933}
K.~P. Bohnen, R.~Heid, B.~Renker, Phys Rev Lett 86 (2001) 5771.

\bibitem{Mittal:2008p13172}
R.~Mittal, Y.~Su, S.~Rols, T.~Chatterji, S.~L. Chaplot, H.~Schober, M.~Rotter,
  D.~Johrendt, T.~Brueckel, Phys Rev B 78 (2008) 104514.

\bibitem{Mittal:2008p15146}
R.~Mittal, Y.~Su, S.~Rols, M.~Tegel, S.~L. Chaplot, H.~Schober, T.~Chatterji,
  D.~Johrendt, T.~Brueckel, Phys Rev B 78 (2008) 224518.

\bibitem{Qiu:2008p12585}
Y.~Qiu, M.~Kofu, W.~Bao, S.-H. Lee, Q.~Huang, T.~Yildirim, J.~R.~D. Copley,
  J.~Lynn, T.~Wu, G.~Wu, X.~H. Chen, Phys Rev B 78 (2008) 052508.

\bibitem{Huang:2008p14968}
Q.~Huang, Y.~Qiu, W.~Bao, M.~A. Green, J.~W. Lynn, Y.~C. Gasparovic, T.~Wu,
  G.~Wu, X.~H. Chen, Phys Rev Lett 101 (2008) 257003.

\bibitem{Yildirim:2008p11683}
T.~Yildirim, Phys Rev Lett 101 (2008) 057010.

\bibitem{Mazin:2008p12399}
I.~I. Mazin, M.~D. Johannes, K.~Koepernik, D.~Singh, Phys Rev B 78 (2008)
  085104.

\bibitem{Ewings:2008p15404}
R.~A. Ewings, T.~G. Perring, R.~I. Bewley, T.~Guidi, M.~J. Pitcher, D.~R.
  Parker, S.~J. Clarke, A.~T. Boothroyd, Phys Rev B 78 (2008) 220501.

\bibitem{Yao:2008p7354}
D.-X. Yao, E.~Carlson, arXiv:0804.4115.

\bibitem{McQueeney:2008p15157}
R.~McQueeney, S.~O. Diallo, V.~P. Antropov, G.~D. Samolyuk, C.~Broholm, N.~Ni,
  S.~Nandi, M.~Yethiraj, J.~L. Zarestky, J.~J. Pulikkotil, A.~Kreyssig, M.~D.
  Lumsden, B.~N. Harmon, P.~C. Canfield, A.~I. Goldman, Phys Rev Lett 101
  (2008) 227205.

\bibitem{Zhao:2008p13418}
J.~Zhao, D.-X. Yao, S.~Li, T.~Hong, Y.~Chen, S.~Chang, W.~Ratcliff, J.~Lynn,
  H.~A. Mook, G.~F. Chen, J.~L. Luo, N.~L. Wang, E.~W. Carlson, J.~Hu, P.~Dai,
  Phys Rev Lett 101 (2008) 167203.

\bibitem{Diallo:2009p15798}
S.~O. Diallo, V.~P. Antropov, C.~Broholm, T.~G. Perring, J.~J. Pulikkotil,
  S.~L. Bud'ko, P.~C. Canfield, A.~Kreyssig, A.~I. Goldman, R.~J. McQueeney,
  arXiv:0901.3784.

\bibitem{Matan:2008p13460}
K.~Matan, R.~Morinaga, K.~Iida, T.~J. Sato, arXiv:0810.4790.

\bibitem{Fawcett:1988p15790}
E.~Fawcett, Rev Mod Phys 60 (1988) 209--283.

\bibitem{Liu:2008p14078}
H.~Liu, W.~Zhang, L.~Zhao, X.~Jia, J.~Meng, G.~Liu, X.~Dong, G.~F. Chen, J.~L.
  Luo, N.~L. Wang, W.~Lu, G.~Wang, Y.~Zhou, Y.~Zhu, X.~Wang, Z.~Xu, C.~Chen,
  X.~J. Zhou, Phys Rev B 78 (2008) 184514.

\bibitem{Ishikado:2008p13337}
M.~Ishikado, R.~Kajimoto, S.~ichi Shamoto, M.~Arai, A.~Iyo, K.~Miyazawa, P.~M.
  Shirage, H.~Kito, H.~Eisaki, S.~Kim, H.~Hosono, T.~Guidi, R.~Bewley,
  S.~Bennington, arXiv:0809.5128.

\bibitem{Cooke:1980p18413}
J.~Cooke, J.~Lynn, H.~Davis, Phys. Rev. B 21 (1980) 4118--4131.

\bibitem{Schrieffer:1964p18081}
J.~R. Schrieffer, \textit{Theory of Superconductivity} (Benjamin, Reading, MA, 1964).

\bibitem{Fong:1995p18080}
H.~F. Fong, B.~Keimer, P.~W. Anderson, F.~Do{\u g}an, I.~A. Aksay, Phys Rev
  Lett 75 (1995) 316--319.

\bibitem{RossatMignod:1991p10509}
J.~Rossat-Mignod, L.~Regnault, C.~Vettier, P.~Bourges, P.~Burlet, J.~Bossy,
  J.~Y. Henry, G.~Lapertot, Physica C 185 (1991) 86.

\bibitem{Mook:1993p10706}
H.~A. Mook, M.~Yethiraj, G.~Aeppli, T.~E. Mason, T.~Armstrong, Phys Rev Lett 70
  (1993) 3490--3493.

\bibitem{Fong:1999p11512}
H.~F. Fong, P.~Bourges, Y.~Sidis, L.~Regnault, A.~Ivanov, G.~D. Gu,
  N.~Koshizuka, B.~Keimer, Nature 398 (1999) 588--591.

\bibitem{Dai:2000p10619}
P.~Dai, H.~A. Mook, G.~Aeppli, S.~Hayden, F.~Do{\u g}an, Nature 406 (2000) 965.

\bibitem{He:2002p11430}
H.~He, P.~Bourges, Y.~Sidis, C.~Ulrich, L.~Regnault, S.~Pailh{\`e}s, N.~S.
  Berzigiarova, N.~N. Kolesnikov, B.~Keimer, Science 295 (2002) 1045--1047.

\bibitem{Ding:2008p8816}
H.~Ding, P.~Richard, K.~Nakayama, K.~Sugawara, T.~Arakane, Y.~Sekiba,
  A.~Takayama, S.~Souma, T.~J. Sato, T.~Takahashi, Z.~Wang, X.~Dai, Z.~Fang,
  G.~F. Chen, J.~L. Luo, N.~Wang, Europhys Lett 83 (2008) 47001.

\bibitem{Korshunov:2008p13468}
M.~M. Korshunov, I.~Eremin, Phys Rev B 78 (2008) 140509.

\bibitem{Maier:2008p11868}
T.~A. Maier, D.~Scalapino, Phys Rev B 78 (2008) 020514.

\bibitem{Inotani:2009p15561}
D.~Inotani, Y.~Ohashi, arXiv:0901.1718.

\end{thebibliography}

\end{document}